# Deep Learning in Detection and Diagnosis of Covid-19 using Radiology Modalities: A Systematic Review

Mustafa Ghaderzadeh[1] , Farkhondeh Asadi[2*]

## Abstraction

**Purpose:** Early detection and diagnosis of Covid-19 and accurate separation of patients with non-Covid-19 cases at the lowest cost and in the early stages of the disease is one of the main challenges in the epidemic of Covid-19. Concerning the novelty of the disease, the diagnostic methods based on radiological images suffer shortcomings despite their many uses in diagnostic centers. Accordingly, medical and computer researchers tended to use machine-learning models to analyze radiology images.

**Methods:** Present systematic review was conducted by searching three databases of PubMed, Scopus, and Web of Science from November 1, 2019, to July 20, 2020 Based on a search strategy, the keywords was "Covid-19, "Deep learning", "Diagnosis", and "Detection" leading to the extraction of 168 articles that ultimately, 37 articles were selected as the research population by applying inclusion and exclusion criteria.

**Result:** This review study provides an overview of the current state of all models for the detection and diagnosis of Covid-19 through radiology modalities and their processing based on deep learning. According to the finding, Deep learning Based models have an extraordinary capacity to achieve an accurate and efficient system for the detection and diagnosis of Covid-19, which using of them in the processing of CT-Scan and X-Ray images, would lead to a significant increase in sensitivity and specificity values.

**Conclusion:** The Application of Deep Learning (DL) in the field of Covid-19 radiologic images processing leads to the reduction of false-positive and negative errors in the detection and diagnosis of this disease and provides an optimal opportunity to provide fast, cheap, and safe diagnostic services to patients.

---

[1] Student Research Committee, Department and Faculty of Health Information Technology and Management, School of Allied Medical Science, Shahid Beheshti University of Medical Sciences, Tehran, Iran

[2] Associate Professor at Department of Health Information Technology and Management, School of Allied Medical Science, Shahid Beheshti University of Medical Sciences, Tehran, Iran (* **Corresponding Author:** Asadifar@sbmu.ac.ir)

# 1. Introduction

With the outbreak of an unknown disease in late 2019 in China, some people became infected with the disease in a local market. Patients were admitted to hospitals with abnormal symptoms, such as fever above 38 degrees, infection, dizziness, muscle aches, and shortness of breath, and cough. Due to respiratory problems and low blood oxygen levels, these people were transferred to intensive care units and many of them lost their lives. The disease was initially completely unknown, but specialists diagnosed the symptoms of the disease as something similar to those of Corona and the flu. The specific cause of this widespread disease was initially unknown, but after laboratory examination and analysis of positive sputum by Real-Time PCR test, viral infection was confirmed and was eventually named Covid-19 by the recommendation of the World Health Organization. This disease was named Covid-2019 with regard to the family of Corona diseases and the year 2019 [1,2] Over a short period, the Covid-19 epidemic crossed geographical boundaries with a devastating effect on the health, economy, and welfare of the global population. According to the reports of the World Health Organization, Covid-19 showed its severity in 16 to 21% of the patients with 2 to 3% of death [3]. Local statistics in many countries indicate a high mortality rate among patients, and some studies have calculated the mortality rate at 4% [4]. Concerning the novelty of the disease, ways to fight it were not known in the early days, but researchers consider screening and rapid diagnosis of infected patients and their separation from the community of healthy people as an important step in fighting against this disease. Clinical features of Covid-19 include respiratory symptoms, fever, cough, shortness of breath, and pneumonia. However, these symptoms do not always indicate Covid-19 and are observed in many cases of pneumonia, leading to diagnostic problems for physicians [5, 6].

According to the latest protocol for the diagnosis and treatment of pneumonia infections caused by Covid-19 (Sixth Edition) published by the Government of China, the main screening method used to diagnose Covid-19 cases is the reverse transcriptional polymerase chain reaction (RT-PCR) test that can distinguish Covid-19 from respiratory diseases and other lung infections [7]. While the RT-PCR test is the gold standard for diagnosing Covid-19, it has limiting aspects with special features that make it difficult to diagnose the diseases. RT-PCR is a very time consuming and complex manual process and Covid-19 diagnosis processes through this method are very costly and time-consuming. The use of diagnostic test requires equipped facilities and laboratories that do not exist even in many developed countries. One of the drawbacks of this method is the need for a laboratory kit, that the provision of which is difficult and impossible for many countries during crises and epidemics. Besides, the provision of diagnostic kits is difficult and impossible for many countries, because it imposes a huge financial burden on their health systems. Like all diagnostic and laboratory methods in health care systems, this method is not error-free and has bias and error. This diagnostic method requires an expert laboratory technician to sample the nasal and throat mucosa which is a painful method and many people refuse to do nasal swap sampling for this reason [8-11]. Most importantly, many studies indicated the low rate of the sensitivity of the RT-PCR test, and a number of studies have reported the sensitivity of this method of diagnosis between 30% to 60%, indicating a decrease in the accuracy of the diagnosis of Covid-19 in many cases. Several study these studies pointed to the false-negative rate and contradictory results of this diagnostic test [11-14].

## 1.1. Detection and Diagnosis of Covid-19 Using Radiology Modality

One of the most important ways to diagnose Covid-19 is to use radiological images. X-ray imaging and CT-Scan are the main screening methods for identification of Covid-19. Radiologists look for visual indicators of lung infection by examining the images. Chest imaging is a quick and easy procedure recommended by medical and health protocols. Chest imaging has been mentioned in several texts as the first tool in screening for disease in epidemics [3, 6, 15].

CT-Scan images compared to RT-PCR have high sensitivity in diagnosing and detecting cases with Covid-19, however, its specificity is less [16]. This means that CT-Scan is more accurate in cases of Covid-19, but less accurate in cases of non-viral pneumonia. In most recorded cases, the sensitivity and specificity of the CT-Scan test in diagnosing Covid-19 are higher than 94% and less than 60%, respectively, which can lead to problems in diagnosing Covid-19. A study conducted on the diagnosis of patients in Wuhan, China showed that consolidation and ground-glass opacities (GGO) were not observed in CT-Scan imaging in 14% of the images, meaning that 14% of the definitive cases of Covid-19 were misdiagnosed as completely healthy and normal based on their CT-Scan tests. Out of 18 patients with Covid-19 who had GGO with Consolidation, only 12 had GGO and as a result, no consolidation and disease were observed. Despite the presence of consolidation without the advent of GGO in many cases, it was difficult and almost impossible to detect Covid-19. All these cases indicated a defect in the diagnosis of Covid-9 using CT-Scan. [16-20].

The American College of Radiology recommends that CT-Scans should not be used as the first line of diagnosis. Problems such as the risk of transmission of the disease while using a CT-Scan device and its high cost can cause serious complications for the patient and health care systems, so it recommends that in the first step if medical imaging is needed, replace the CT-Scan with CXR radiography [21]. One of the methods of imaging diseases is CXR. Various studies have indicated the failure of CXR imaging to diagnosing Covid-19 and differentiating it from other types of pneumonia. The radiologist cannot use X-rays to detect pleural effusion and determine the volume involved [20]. However, regardless of the low accuracy of X-ray diagnosing Covid-19, it has some positive points:

X-ray imaging is much more extensive and cost-effective than conventional diagnostic tests. Transmission of an X-ray digital image does not require transferring from the access point to the analysis point, so it performs the diagnostic process very quickly. Chest radiography is convenient and fast for medical triage of patients [20]. Unlike CT-Scans, X-ray imaging requires less additional Scarce and valuable equipment, so significant savings can be made in the running costs. Furthermore, portable CXR devices can be used in isolated rooms, to reduce the risk of infection resulted from the use of these devices in hospitals [13, 20].

To overcome the limitations of Covid-19 diagnostic tests using radiological images, various studies have been performed on the use of machine learning and computer-Based methods in the analysis of radiological images. Due to the existing problems in diagnosis and interpreting of radiological images of Covid-19, numerous studies have focused on the use of artificial intelligence and machine learning in the interpretation of radiological images. Machine learning methods, although successful in processing disease data, have their drawbacks in diagnosing diseases through images. One of these drawbacks was the selection and extraction of features by users and their suggested methods. In which a number of features were ignoring or many effective features were in the detection process less involving.

Artificial Intelligence includes random forest methods, support vector machine, nearest neighborhood, and artificial neural network, all of which use computational methods and artificial intelligence to select the features.

In the case of images dealing with large volumes of data, the usual methods of artificial intelligence do not have the necessary efficiency because the input of images is applied to the pixel by pixel, an increase in the number of pixels, its parameters significantly increase, which would impose problems, such as over-fitting, pre-processing and reduction of networks speed in processing [22, 23]. Therefore, traditional machine learning methods were not efficient in diagnosing Covid-19 by radiological images.

## *1.2. Deep Learning*

In 2006, Hinton and Salkhudinoy published an article in Science Journal that was a gateway to the age of deep learning. Hinton showed that the neural network with hidden layers played a great role in increasing the learning power of features. Therefore, these algorithms are able to increase the accuracy of classifying different types of data. Therefore, this field of artificial intelligence has achieved extraordinary success in medical image processing [23, 24]. Deep learning covered the disadvantages of conventional machine learning methods and revolutionized the selection and extraction of features. One of the major applications of deep learning in radiology practices was the detection of tissue-skeletal abnormalities and the classification of diseases. The Convolutional neural network has proven to be one of the most important deep learning algorithms as the most effective technique in detecting abnormalities and pathology of chest radiographs [25-27]. Since the outbreak of Covid-19, much research has been conducted on processing the related data to deep learning algorithms, especially CNN. Using different algorithms and deep learning architectures, these studies have embarked on the identification and differential diagnosis of Covid-19. In this study, these studies have been systematically analyzed.

# *2. Methodology*

This study was accomplished by a structured review method to identify studies related to the identification and diagnosis of Covid-19. A Systematic search strategy was performed by using previous studies and the author's opinions.

## *2.1. Search Criteria*

1) To what extent has the use of deep learning been able to improve the usual methods of diagnosing Covid-19?
2) What modalities can be used to help identify and diagnose Covid-19 by using deep learning?
3) Has deep learning been able to cover the shortcomings of diagnostic modalities?
4) What is the efficacy of different types of deep learning and its architectures in promoting the diagnosis of Covid-19 compared to each other?

The researchers reviewed the electronic databases to identify studies on medicine and computer sciences and concluded that PubMed, Web of Science, and Scopus databases have the highest number of publications related to the current study. The following key terms were used as the search strategy: "Covid-19", "diagnosis", "detection" and "deep learning" from November 1, 2019, to July 20, 2020, and related published studies were extracted from the three databases. The EMBASE and IEEE databases were removed from the search domain due to the proximity of their publications.

## *2.2. Data extraction*

Related studies, details of their methodologies, and their results were recorded in data extraction forms. Data selection and extraction in the present study were performed based on the Figure (1). To identify algorithms and deep learning methods, the main details of the methods and their results were recorded in data extraction sheets. Two researchers (FA)[1] and (M.G)[2] extracted data and differences between studies were resolved through discussion. The extracted data elements included the name of the study, country, year of publication, research population, modality, or the data used deep learning techniques, and evaluation methods and results.

## 3. Result

Initially, a total of 160 abstracts and full-text articles were assessed, and ultimately 37 studies meeting the inclusion criteria were selected. PRISMA method was used in the process of selecting the articles. Due to the novelty of the disease, all selected articles have been published in 2020. Out of 37 extracted articles, 8 articles were published in India, 5 in China, 5 in the United States, and 3 in Turkey. Also, Iran, Greece, Italy, and Egypt each presented two research papers, and Morocco, Bangladesh, Spain, Colombia, Iraq, Brazil, Canada, and South Korea each presented one study in this field.

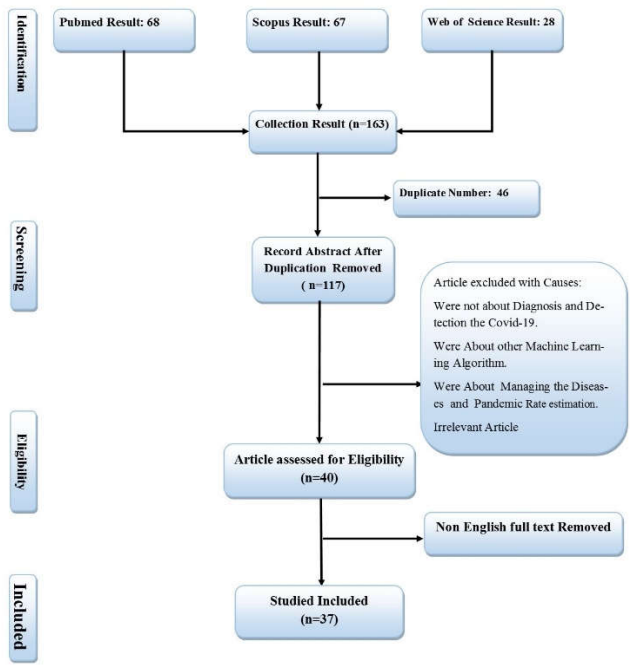

*Figure 1: PRISMA flow diagram of study review process and exclusion of papers*

## *3.1) Purpose of Deep learning in analysis Radiology Images about Covid-19*

Image-based diagnostic methods in epidemics play a key role in the screening of affected cases. CXR and CT-Scan are among the main radiology modalities in the detection and diagnosis of Covid-19. In all the studies reviewed in this study, radiological images have been analyzed to diagnose Covid-19 with deep learning. This study has been conducted around two common terms, i.e. "detection" and "diagnosis" to characterize the presence of Covid-19. Collins Dictionary defines the terms "detection" and "diagnosis" differently from the medical point of view. However, in the field of medical image analysis, these two terms have been used repeatedly for the same purpose. By examining the existing texts and dictionaries and seeking advice from radiologists and epidemiologists, detection is part of the real entity that can be seen or its existence can be proved or disproved.

In medical texts, detection is considered as a prelude to diagnosis. Similarly, in the case of Covid-19, many studies have used these two words interchangeably, which are clinically different from each other. By distinguishing between these two terms in this study, detection has been considered as the distinguishing cases infected with Covid-19 from non-Covid-19. This means that no information is available on the type of the disease in non-Covid-19 patients, and this group can have different types of bacterial pneumonia, viral, or other groups of coronavirus diseases with the exception of Covid-19. We also considered diagnosis as a term to distinguish Covid-19 from other infectious lung diseases such as different types of pneumonia. Diagnosis is meaningful in categories where the rest of the diseases, not infected with Covid-19, are well specified, and Covid-19 can be distinguished with certainty from types of pneumonia or other coronaviruses. In this regard, by examining the articles extracted in this study, it was found that 15 articles had used deep learning to detect (identify) Covid-19 [28-42]. On the other hand, many articles have diagnosed Covid-19 with deep learning algorithms [43-59]. In these cases, Covid-19 was accurately diagnosed among the different types of pneumonia. Some studies have analyzed radiology modalities to detect and diagnose it simultaneously [39, 60]. Figure (2) shows studies on the detection and diagnosis of Covid-19.

---

[1] Second author
[2] First author

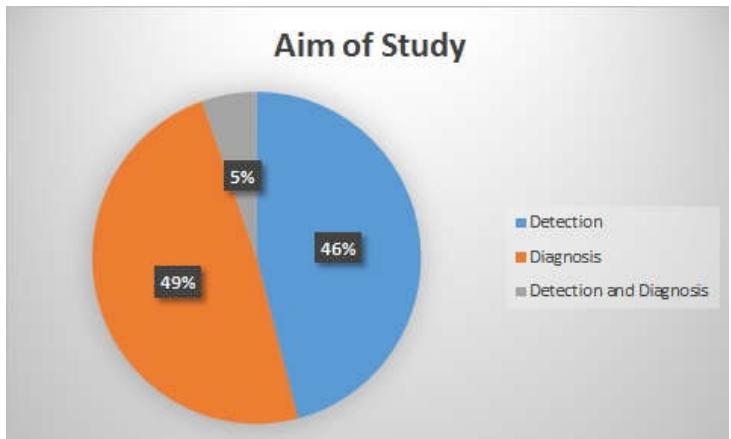

*Figure 2: Research methods for detection and diagnosis of Covid-19 by means of deep learning*

It has already been said that one of the diagnostic disadvantages of CT-Scan images in identification cases of Covid-19 its low specificity. Accordingly, the investigation has shown many studies conducted to improve these methods in the analysis of CT-Scan images with deep learning techniques [28, 29, 33-35, 40, 46, 49, 50, 52, 61-65]. Apparently, these methods owe their success in finding pulmonary lesion caused by Covid-19 to the extraction and selection of features hidden in the images. Despite the improvement in the detection and diagnosis of Covid-19 by deep learning algorithms, one of the biggest drawbacks of this modality in the diagnosis of Covid-19 was the lack of this equipment in all medical and diagnostic centers. Furthermore, many patients with Covid-19 required multiple chest images using CT-Scans. Exposure to radiation during CT-Scans causes serious problems for patients Moreover, there is a danger of transmission of the Covid-19 virus from one patient to other people due to CT-Scan tunnel contamination.

Therefore, many researchers and physicians have resorted to plain radiographic images or X-rays to diagnose Covid-19. However, these images do not have the resolution and accuracy in diagnosing Covid-19 from the beginning and have many disadvantages in this regard. Therefore, artificial intelligence researchers rushed to the help of clinical experts and used deep learning as a powerful tool to improve the accuracy of the diagnosis of Covdi-19 with X-ray images [22]. Due to the nature of deep learning in the extraction of image features, this technology is capable of detecting patients with Covid-19 and extracting infectious lung tissue, so many studies embarked on a variety of deep learning algorithms to analyze these images [30-32, 36-39, 41, 42, 53, 58-60, 66-71]. In the early days of the Covid-19 outbreak, CT-Scans were more common in diagnosing it, but over time, X-rays also became common. Therefore, research also shifted from CT-Scan image analysis to radiographic image analysis. Figure (3) shows the amount of analysis of the two modalities used to detect and diagnose Covid-19.

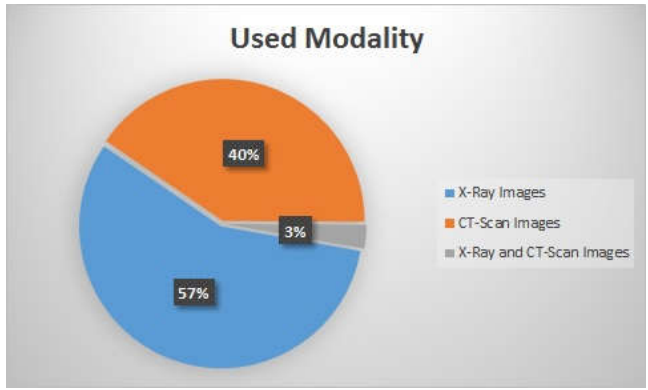

Figure 3: rate of using different radiological modalities in processing of Covid-19 with deep learning

One of the main features of deep neural networks, in terms of its efficacy, is its employed architecture. Deep neural network architectures demonstrate an extraordinary ability to perform a variety of functions for a variety of data types. Different studies have been performed on Covid-19 with different deep learning architectures. In a number of these studies, their diagnosis rate was compared in the detection of Covid-19 by using different types of architectures [28, 33]. The frequency of CNN architectures used in the present study can be seen in Figure (4). The architectures presented in this figure represent a family of the same architectures or different editions of that architecture. In the texts reviewed, the ResNet architecture enjoyed the most efficacy. However, some of the studies with ResNet 50 architecture achieved the best efficacy in detecting and diagnosing Covid-19, and others used other ResNet editions to maximize efficacy in analyzing radiological images for the diagnosis of Covid-19. It was found that newer and more developed architectures were more efficient in diagnosing Covid-19.

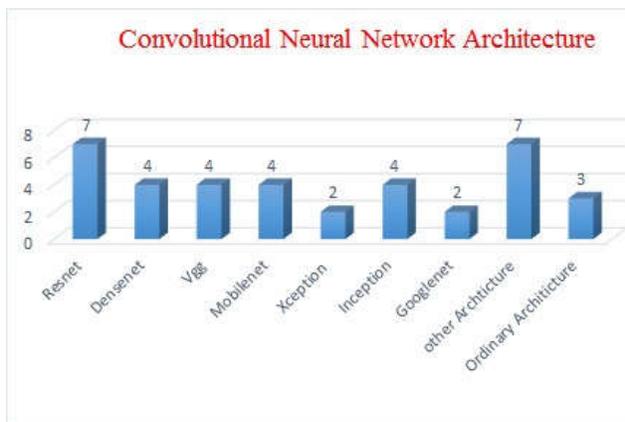

Figure 4: rate of using radiological modalities in processing Covid-19 with deep learning.

## 3.2 Deep learning techniques in Detection and Diagnosis of Covid-19

Studies suggest that different deep learning techniques have been used to detect, diagnose, classify, predict, and prognosis of Covid-19. Domestic datasets (including CT and X-ray images) or public datasets have been used in some studies in which. training and testing datasets were used to teach and validate the methods. Criteria for measuring the efficiency of methods used in community studies include sensitivity, specificity, and accuracy. However, in many research, AUC has also been used to determine the efficiency of the method used to diagnose Covid-19.

In a number of studies, the proposed method has been implemented based on well-known or state of the art architectures. However, some studies have also presented their own customized algorithm and architecture, independent of well-known architectures.

Studies suggest that different deep learning techniques have been used to detect, diagnose, classify, predict, and prognosis of Covid-19. Domestic datasets (including CT and X-ray images) or public datasets have been used in some studies in which. training and testing datasets were used to teach and validate the methods. Criteria for measuring the efficiency of methods used in community studies include sensitivity, specificity, and accuracy. However, in some studies, AUC has also been used to determine the efficiency of the method used to diagnose Covid-19. In a number of studies, the proposed method has been implemented based on well-known or state of the art architectures. However in many studies have also presented their own customized algorithm and architecture, independent of well-known architectures.

## *4: Discussion*

This systematic study evaluated 37 studies that work to assist researchers to explore and develop knowledge-based systems based on artificial intelligence in the detection and diagnosis of Covid-19. To improve our knowledge, the current review, which uses a variety of deep learning methods to analyze radiological images, is one of the most comprehensive studies in diagnosis and detection Covid-19. The current review provided up-to-date information on deep learning algorithms and their application as an expression of radiographic imaging analysis of Covid-19. Many studies have shown that the use of deep learning algorithms can improve the rate of metric features of CT-Scan images and increase the sensitivity and specificity of radiographic images compared to the radiologist's diagnosis; therefore, the use of this cheap and affordable modality should be considered as a reliable method for the diagnosis of Covid-9. By reviewing 23 research papers on the application of X-ray in the diagnosis of Covid-19 by using deep learning methods, the current modality can be introduced to the scientific and medical community in the field of early and faster diagnosis of Covid-19. By improving imaging methods through artificial intelligence technologies, we can point to the cheapest and safest imaging methods to prevent the transmission of Covid-19. A review of published studies showed that the diagnosis of Covid-19 by deep learning algorithms under the supervision of a radiologist led to an improved efficacy and reduced diagnostic errors in various cases of pneumonia, especially Covid-19. Mean diagnosis of all the studies performed through X-ray modality had a sensitivity with an average higher than 95% and a specificity of higher than 91%, which had a higher rate of diagnosis than that of Covid-19 in traditional texts and methods.

It can also be concluded from the studies that the specificity in CT-Scan images obtained by deep learning method in case of contracting Covid-19 was on average higher than 92%, which in many cases, compared to previous texts, has a higher efficiency in terms of the specificity; the sensitivity of deep learning methods in CT-Scan images of Covid-19 was also higher than or equal to the usual diagnostic methods in many cases. It is thought that due to the over-similarity of the effects of Covid-19 on lung tissue with different types of bacterial and viral pneumonia, the diagnosis of these diseases through unsupervised methods is very difficult and complicated. The examination of the algorithms and deep learning architectures revealed that almost all studies have used the CNN algorithm; of course, other algorithms have been used along with the CNN algorithm in other studies. The CNN architectures used in these studies all have special features in image analysis, and without adjusting their parameters, it is not possible to have access to the ability of these architectures to detect and diagnose Covid-19.

## *5. Conclusion and Future work*

As discussed in this review research, the early detection and diagnosis of Covid-19, by deep learning techniques and with the least cost and complications are the basic steps in preventing the disease and the progression of the pandemic. In the near future, with the incorporation of deep learning algorithms in the equipment of radiology centers, it will be possible to achieve a faster, cheaper, and safer diagnosis of Covid-19. The use of these techniques in rapid diagnostic decision-making of Covid-19 can be a powerful tool for a radiologist to reduce human error and assist them to make decisions in critical conditions and the peak of the disease. This research supports the idea that deep learning algorithms are a promising way of optimizing health care and improving the results of diagnostic and therapeutic procedures. Although deep learning is one of the most powerful computing tools in diagnosis pneumonia, especially Covid-19, developers should be careful to avoid overfitting and to maximize the generalizability and usefulness of Covid-19 deep learning diagnostic models; these models must be trained on large, heterogeneous datasets to cover all available data space.

# 6. Acknowledgement

This study is related to the project NO. …………........................... From Student Research Committee, Shahid Beheshti University of Medical Sciences,Tehran,Iran. We also appreciate the "Student Research Committee" and "Research & Technology Chancellor" in Shahid Beheshti University of Medical Sciences for their financial support of this study.

1. Bravo, D., et al., *Effect of the IL28B Rs12979860 C/T polymorphism on the incidence and features of active cytomegalovirus infection in allogeneic stem cell transplant patients.* Journal of Medical Virology, 2014. **86**(5): p. 838-844.
2. Chen, N., et al., *Epidemiological and clinical characteristics of 99 cases of 2019 novel coronavirus pneumonia in Wuhan, China: a descriptive study.* The Lancet, 2020. **395**(10223): p. 507-513.
3. Wang, L. and A. Wong, *COVID-Net: A Tailored Deep Convolutional Neural Network Design for Detection of COVID-19 Cases from Chest X-Ray Images.* arXiv preprint arXiv:2003.09871, 2020.
4. Baud, D., et al., *Real estimates of mortality following COVID-19 infection.* The Lancet infectious diseases, 2020.
5. Li, Q., et al., *Early transmission dynamics in Wuhan, China, of novel coronavirus–infected pneumonia.* New England Journal of Medicine, 2020.
6. Huang, C., et al., *Clinical features of patients infected with 2019 novel coronavirus in Wuhan, China.* The lancet, 2020. **395**(10223): p. 497-506.
7. Committee, G.O.o.N.H., *Notice on the issuance of a program for the diagnosis and treatment of novel coronavirus (2019-nCoV) infected pneumonia (trial revised fifth edition).* 2020.
8. Lippi, G. and M. Plebani, *A Six-Sigma approach for comparing diagnostic errors in healthcare—where does laboratory medicine stand?* Annals of translational medicine, 2018. **6**(10).
9. Sheridan, C., *Coronavirus and the race to distribute reliable diagnostics.* Nature biotechnology, 2020. **38**(4): p. 382.
10. Lippi, G., A.-M. Simundic, and M. Plebani, *Potential preanalytical and analytical vulnerabilities in the laboratory diagnosis of coronavirus disease 2019 (COVID-19).* Clinical Chemistry and Laboratory Medicine (CCLM), 2020. **1**(ahead-of-print).
11. Oliveira, B.A., et al., *SARS-CoV-2 and the COVID-19 disease: a mini review on diagnostic methods.* Revista do Instituto de Medicina Tropical de Sao Paulo, 2020. **62**.
12. Ai, T., et al., *Correlation of chest CT and RT-PCR testing in coronavirus disease 2019 (COVID-19) in China: a report of 1014 cases.* Radiology, 2020: p. 200642.
13. Fang, Y., et al., *Sensitivity of chest CT for COVID-19: comparison to RT-PCR [published online February 19, 2020].* Radiology. doi. **10**.
14. Wang, Y., et al., *Unique epidemiological and clinical features of the emerging 2019 novel coronavirus pneumonia (COVID-19) implicate special control measures.* Journal of medical virology, 2020. **92**(6): p. 568-576.
15. Wang, W., et al., *Detection of SARS-CoV-2 in different types of clinical specimens.* Jama, 2020. **323**(18): p. 1843-1844.
16. Gozes, O., et al., *Rapid ai development cycle for the coronavirus (covid-19) pandemic: Initial results for automated detection & patient monitoring using deep learning ct image analysis.* arXiv preprint arXiv:2003.05037, 2020.
17. Wang, S., et al., *A deep learning algorithm using CT images to screen for Corona Virus Disease (COVID-19).* MedRxiv, 2020.

18. Salehi, S., et al., *Coronavirus disease 2019 (COVID-19): a systematic review of imaging findings in 919 patients.* American Journal of Roentgenology, 2020: p. 1-7.
19. Zhao, J., et al., *COVID-CT-Dataset: a CT-Scan dataset about COVID-19.* arXiv preprint arXiv:2003.13865, 2020.
20. Rubin, G.D., et al., *The role of chest imaging in patient management during the COVID-19 pandemic: a multinational consensus statement from the Fleischner Society.* Chest, 2020.
21. Radiology, A.C.o., *ACR recommendations for the use of chest radiography and computed tomography (CT) for suspected COVID-19 infection.* ACR website. Advocacy-and Economics/ACR-Position-Statements/Recommendations-for-Chest-Radiography-and-CTfor-Suspected-COVID-19-Infection. Updated March, 2020. **22**.
22. Elaziz, M.A., et al., *New machine learning method for image-based diagnosis of COVID-19.* Plos one, 2020. **15**(6): p. e0235187.
23. Suzuki, K., *Machine Learning in Computer-Aided Diagnosis: Medical Imaging Intelligence and Analysis: Medical Imaging Intelligence and Analysis*. 2012: IGI Global.
24. Ker, J., et al., *Deep learning applications in medical image analysis.* Ieee Access, 2017. **6**: p. 9375-9389.
25. Wang, X., et al. *Hospital-scale chest x-ray database and benchmarks on weakly-supervised classification and localization of common thorax diseases*. in *IEEE CVPR*. 2017.
26. Lu, M.T., et al., *Deep learning to assess long-term mortality from chest radiographs.* JAMA network open, 2019. **2**(7): p. e197416-e197416.
27. Kieu, P.N., et al. *Applying multi-CNNs model for detecting abnormal problem on chest x-ray images*. in *2018 10th International Conference on Knowledge and Systems Engineering (KSE)*. 2018. IEEE.
28. Ardakani, A.A., et al., *Application of deep learning technique to manage COVID-19 in routine clinical practice using CT images: Results of 10 convolutional neural networks.* Computers in Biology and Medicine, 2020. **121**.
29. Ni, Q., et al., *A deep learning approach to characterize 2019 coronavirus disease (COVID-19) pneumonia in chest CT images.* European Radiology, 2020.
30. Panwar, H., et al., *Application of deep learning for fast detection of COVID-19 in X-Rays using nCOVnet.* Chaos, Solitons and Fractals, 2020. **138**.
31. Saiz, F.A. and I. Barandiaran, *COVID-19 Detection in Chest X-ray Images using a Deep Learning Approach.* International Journal of Interactive Multimedia and Artificial Intelligence, 2020. **6**(2): p. 11-14.
32. Vaid, S., R. Kalantar, and M. Bhandari, *Deep learning COVID-19 detection bias: accuracy through artificial intelligence.* International Orthopaedics, 2020.
33. El Asnaoui, K. and Y. Chawki, *Using X-ray images and deep learning for automated detection of coronavirus disease.* Journal of Biomolecular Structure & Dynamics.
34. Yang, S., et al., *Deep learning for detecting corona virus disease 2019 (COVID-19) on high-resolution computed tomography: a pilot study.* Ann Transl Med, 2020. **8**(7): p. 450.
35. Jaiswal, A., et al., *Classification of the COVID-19 infected patients using DenseNet201 based deep transfer learning.* Journal of Biomolecular Structure and Dynamics, 2020: p. 1-8.
36. Apostolopoulos, I.D. and T.A. Mpesiana, *Covid-19: automatic detection from X-ray images utilizing transfer learning with convolutional neural networks.* Physical and Engineering Sciences in Medicine, 2020. **43**(2): p. 635-640.
37. Ozturk, T., et al., *Automated detection of COVID-19 cases using deep neural networks with X-ray images.* Computers in Biology and Medicine, 2020. **121**.
38. Waheed, A., et al., *CovidGAN: Data Augmentation Using Auxiliary Classifier GAN for Improved Covid-19 Detection.* Ieee Access, 2020. **8**: p. 91916-91923.
39. Brunese, L., et al., *Explainable Deep Learning for Pulmonary Disease and Coronavirus COVID-19 Detection from X-rays.* Computer Methods and Programs in Biomedicine, 2020. **196**.

40. Song, J., et al., *End-to-end automatic differentiation of the coronavirus disease 2019 (COVID-19) from viral pneumonia based on chest CT.* European journal of nuclear medicine and molecular imaging, 2020: p. 1-9.
41. Martínez, F.e.a., *Performance Evaluation of the NASNet Convolutional Network in the Automatic Identification of COVID-19.* International Journal on Advanced Science, Engineering and Information Technology, 2020. **10**(2).
42. Yi, P.H., T.K. Kim, and C.T. Lin, *Generalizability of Deep Learning Tuberculosis Classifier to COVID-19 Chest Radiographs: New Tricks for an Old Algorithm?* Journal of Thoracic Imaging, 2020.
43. Ko, H., et al., *COVID-19 Pneumonia Diagnosis Using a Simple 2D Deep Learning Framework With a Single Chest CT Image: Model Development and Validation.* Journal of medical Internet research, 2020. **22**(6): p. e19569.
44. Wang, Y.C., et al., *Dynamic evolution of COVID-19 on chest computed tomography: experience from Jiangsu Province of China.* European Radiology.
45. Rahimzadeh, M. and A. Attar, *A modified deep convolutional neural network for detecting COVID-19 and pneumonia from chest X-ray images based on the concatenation of Xception and ResNet50V2.* Informatics in Medicine Unlocked, 2020. **19**.
46. Li, L., et al., *Artificial Intelligence Distinguishes COVID-19 from Community Acquired Pneumonia on Chest CT.* Radiology, 2020: p. 200905.
47. Apostolopoulos, I.D., S.I. Aznaouridis, and M.A. Tzani, *Extracting Possibly Representative COVID-19 Biomarkers from X-ray Images with Deep Learning Approach and Image Data Related to Pulmonary Diseases.* Journal of Medical and Biological Engineering, 2020. **40**(3): p. 462-469.
48. Toğaçar, M., B. Ergen, and Z. Cömert, *COVID-19 detection using deep learning models to exploit Social Mimic Optimization and structured chest X-ray images using fuzzy color and stacking approaches.* Comput Biol Med, 2020. **121**: p. 103805.
49. Wu, X., et al., *Deep learning-based multi-view fusion model for screening 2019 novel coronavirus pneumonia: A multicentre study.* European Journal of Radiology, 2020. **128**.
50. Mei, X., et al., *Artificial intelligence–enabled rapid diagnosis of patients with COVID-19.* Nature Medicine, 2020: p. 1-5.
51. Pereira, R.M., et al., *COVID-19 identification in chest X-ray images on flat and hierarchical classification scenarios.* Computer Methods and Programs in Biomedicine, 2020. **194**.
52. Hasan, A.M., et al., *Classification of Covid-19 Coronavirus, Pneumonia and Healthy Lungs in CT-Scans Using Q-Deformed Entropy and Deep Learning Features.* Entropy, 2020. **22**(5).
53. Loey, M., F. Smarandache, and N.E.M Khalifa, *Within the Lack of Chest COVID-19 X-ray Dataset: A Novel Detection Model Based on GAN and Deep Transfer Learning.* Symmetry-Basel, 2020. **12**(4).
54. Ucar, F. and D. Korkmaz, *COVIDiagnosis-Net: Deep Bayes-SqueezeNet based diagnosis of the coronavirus disease 2019 (COVID-19) from X-ray images.* Medical Hypotheses, 2020. **140**.
55. Singh, K.K., M. Siddhartha, and A. Singh, *Diagnosis of Coronavirus Disease (COVID-19) from Chest X-Ray images using modified XceptionNet.* Romanian Journal of Information Science and Technology, 2020. **23**: p. S91-S105.
56. Mahmud, T., M.A. Rahman, and S.A. Fattah, *CovXNet: A multi-dilation convolutional neural network for automatic COVID-19 and other pneumonia detection from chest X-ray images with transferable multi-receptive feature optimization.* Computers in Biology and Medicine, 2020. **122**.
57. Butt, C., et al., *Deep learning system to screen coronavirus disease 2019 pneumonia.* Applied Intelligence.
58. Sethy, P.K., et al., *Detection of coronavirus disease (COVID-19) based on deep features and support vector machine.* International Journal of Mathematical, Engineering and Management Sciences, 2020. **5**(4): p. 643-651.

59. Das, D., K.C. Santosh, and U. Pal, *Truncated inception net: COVID-19 outbreak screening using chest X-rays.* Physical and Engineering Sciences in Medicine.
60. Khan, A.I., J.L. Shah, and M.M. Bhat, *Coronet: A deep neural network for detection and diagnosis of COVID-19 from chest x-ray images.* Computer Methods and Programs in Biomedicine, 2020: p. 105581.
61. Ramalingam, B., et al., *A human support robot for the cleaning and maintenance of door handles using a deep-learning framework.* Sensors (Switzerland), 2020. **20**(12): p. 1-18.
62. Wang, S., et al., *A Fully Automatic Deep Learning System for COVID-19 Diagnostic and Prognostic Analysis.* Eur Respir J, 2020.
63. Singh, D., et al., *Classification of COVID-19 patients from chest CT images using multi-objective differential evolution-based convolutional neural networks.* Eur J Clin Microbiol Infect Dis, 2020. **39**(7): p. 1379-1389.
64. Pathak, Y., et al., *Deep Transfer Learning Based Classification Model for COVID-19 Disease.* IRBM, 2020.
65. Butt, C., et al., *Deep learning system to screen coronavirus disease 2019 pneumonia.* Applied Intelligence, 2020.
66. Apostolopoulos, I.D., S.I. Aznaouridis, and M.A. Tzani, *Extracting possibly representative COVID-19 Biomarkers from X-Ray images with Deep Learning approach and image data related to Pulmonary Diseases.* Journal of Medical and Biological Engineering, 2020: p. 1.
67. Toğaçar, M., B. Ergen, and Z. Cömert, *COVID-19 detection using deep learning models to exploit Social Mimic Optimization and structured chest X-ray images using fuzzy color and stacking approaches.* Computers in Biology and Medicine, 2020. **121**.
68. Rajaraman, S. and S. Antani, *Training deep learning algorithms with weakly labeled pneumonia chest X-ray data for COVID-19 detection.* medRxiv, 2020.
69. Brunese, L., et al., *Explainable deep learning for pulmonary disease and coronavirus COVID-19 detection from X-rays.* Computer Methods and Programs in Biomedicine, 2020: p. 105608.
70. Ucar, F. and D. Korkmaz, *COVIDiagnosis-Net: Deep Bayes-SqueezeNet based diagnosis of the coronavirus disease 2019 (COVID-19) from X-ray images.* Med Hypotheses, 2020. **140**: p. 109761.
71. Mahmud, T., M.A. Rahman, and S.A. Fattah, *CovXNet: A multi-dilation convolutional neural network for automatic COVID-19 and other pneumonia detection from chest X-ray images with transferable multi-receptive feature optimization.* Comput Biol Med, 2020. **122**: p. 103869.